\documentclass[prb,twocolumn,showpacs,showkeys,superscriptaddress]{revtex4}
\usepackage{graphicx}

\begin{document}
\title{Concentration dependence of superconductivity and 
order-disorder transition in the hexagonal rubidium tungsten bronze 
Rb$_x$WO$_3$. Interfacial and bulk properties}
\author{R. Brusetti}
\affiliation{Centre de Recherches sur les Tr\`es Basses Temp\'eratures, 
associ\'e \`a l'Universit\'e Joseph Fourier, CNRS, BP~166, 
38042 \mbox{Grenoble}~Cedex~9, France} 
\author{P. Haen}
\affiliation{Centre de Recherches sur les Tr\`es Basses Temp\'eratures, 
associ\'e \`a l'Universit\'e Joseph Fourier, CNRS, BP~166, 
38042 \mbox{Grenoble}~Cedex~9, France}
\author{J. Marcus}
\affiliation{Laboratoire d'Etudes des Propri\'et\'es Electroniques des Solides, 
CNRS, BP~166, 38042~Grenoble~Cedex~9, France}

\begin{abstract}
We revisited the problem of the stability of the 
superconducting state in Rb$_x$WO$_3$ and identified the main causes of the 
contradictory data previously published. We have shown that the 
ordering of the Rb vacancies in the non-stoichiometric compounds have 
a major detrimental effect on the superconducting temperature $T_{\rm c}$. The 
order-disorder transition is first order only near $x$~= 0.25, where it 
cannot be quenched effectively and $T_{\rm c}$ is reduced below 1~K. We found 
that the high $T_{\rm c}$'s, which were sometimes deduced from resistivity 
measurements and attributed to compounds with $0.25\lesssim x\lesssim 0.30$, are 
to be ascribed to interfacial superconductivity which generates 
spectacular non-linear effects. We also clarified the effect of acid 
etching and set more precisely the low-rubidium-content boundary of 
the hexagonal phase. This work makes clear that $T_{\rm c}$ would increase 
continuously (from $\approx 2$~K up to $\approx 5.5$~K) as we approach 
this boundary 
($x\approx 0.20$), if no ordering would take placeÑ--as it is approximately 
the case in Cs$_x$WO$_3$. This behavior is reminiscent of the tetragonal 
tungsten bronze Na$_x$WO$_3$ and asks the same question~: what mechanism is 
responsible for this large increase of $T_{\rm c}$ despite the considerable 
associated reduction of the electron density of state $\cal{D}_{\rm 
FE}$~? By 
reviewing the other available data on these bronzes we conclude that 
the theoretical models which are able to answer this question are 
probably those where the instability of the lattice plays a major 
role and, particularly, the model which call upon local structural 
excitations (LSE), associated with the missing alkali atoms.
\end{abstract}
\maketitle

\section{\label{intro}Introduction}

Tungsten bronzes with general formula M$x$WO$_3$ are the first oxides
where superconductivity has been observed in 1964.\cite{RaubPRL64} 
Much attention has been paid to these non-stoichiometric compounds in
which it seemed possible to study how the normal and superconducting
states react to the filling of the conduction band---as this can be
adjusted within rather large ranges by varying the M content (M~=
alkali).  The most extensive investigations have been carried out on
Na$_x$WO$_3$---which displays distorted perovskite structures---and
particularly on the tetragonal phases, in which a large increase of
the superconducting transition temperature $T_{\rm c}$ has been observed by
reducing the Na content ($T_c\approx 0.7$~K for $x$~= 0.4 whereas
$T_c\approx 3$~K for $x$~= 0.2).\cite{ShanksSSC74} This $T_c(x)$
dependence was rather puzzling because it seemed quite clear that the
density of state at the Fermi level $\cal{D}_{\rm FE}$ was decreasing
with $x$.  At the same time, an enormous amount of data was
accumulated on those properties which are related to the high mobility
of the small M atoms, and which could lead to applications in
electrochromic devices and solid electrolytes.  These highly mobile M
atoms---and the associated Einstein-like phonon modes---were also
suspected to play a determining role in the electron-phonon coupling;
however, no clear correlation has been found between the
characteristic energy of these modes and the stability of the
superconducting state. Many studies have been also devoted to the
hexagonal tungsten bronzes (HTB), where higher $T_{\rm c}$'s were obtained, 
\cite{SweedlerStanley}
but they yielded many conflicting results, particularly about the
$T_c(x)$ dependence, and this apparently discouraged further attempts to
get a better knowledge of these systems.  Recently, however, this
issue came back in the foreground incidentally after superconductivity
and even high-temperature (superficial) superconductivity was observed
in WO$_{3-x}$\cite{AirdJPL98} and Na$_{0.05}$WO$_3$\cite{ReichEPJB99} 
respectively.  This encouraged us to revisit the HTB, and particularly 
Rb$_x$WO$_3$ where the highest $T_{\rm c}$ had been observed.

The structure of these bronzes has been described first by 
Magn\'eli.\cite{MagneliACS53} 
It is based on a framework of WO$_6$ (distorted) octahedra which are
linked by their corners, as they are in the tetragonal tungsten 
bronzes (TTB), but forming here
hexagonal tunnels in which the alkali atoms are accommodated.  This
structure is stabilized by these atoms, if they are large enough (K,
Rb, Cs) and if they fill more than a half of the tunnel sites 
($0.19\lesssim x\lesssim 0.33$).

The electronic structure of the tungsten bronzes has been calculated
only for cubic MWO$_3$ and hexagonal WO$_3$ model 
systems\cite{HjelmPRB96} and is in rather good agreement with the 
results of photoemission measurements carried out on Na$_x$WO$_3$.
\cite{HochstPRB82} These calculations 
seem to indicate that
the main features of the valence and conduction bands are not very
sensitive to the symmetry of the WO$_3$ framework and rather independent
of M: the role of the alkali atoms being above all to give their $s$
electrons to the conduction band whose bottom is mainly built of W
$5d-t_{2g}$ orbitals, hybridized with the O $2p$'s.

The hexagonal tunnels running along the $c$
direction are quite open and this allows the smaller alkali atoms to
be very mobile.  In the substoichiometric compounds Rb$_x$WO$_3$ and 
K$_x$WO$_3$
they tend to order below or near room temperature respectively.
\cite{SatoPRB82,KrauseSSC88} 
According to Sato \textit{et al.}\cite{SatoPRB82} this ordering could 
be responsible for the
drastic reduction of $T_{\rm c}$ around $x\approx 0.25$ and for the 
electron-transport anomalies observed by Stanley \textit{et al.}
\cite{StanleyPRB79} and Cadwell \textit{et al.}\cite{CadwellPRB81}. Similar 
but much less pronounced behaviors were recognized in Cs$_x$WO$_3$ by Skokan 
\textit{et al.},\cite{SkokanPRB79} as if the larger size of the Cs 
atoms impeded their
ordering. These authors---belonging to what we shall refer to as the
Florida group---also reported an unexplained anisotropy of the upper
critical field and Stanley \textit{et al.}\cite{StanleyPRB79} underlined 
the poor reproducibility
of their data: some Rb$_x$WO$_3$ samples displaying large $T_{\rm c}$ 
($\approx 7.5$~K)
whereas no superconductivity was observed in others belonging to the
same composition range. Moreover, other investigations detected no
anomaly in the transport properties.\cite{AristimunoJSSC80} Another 
point was left
controversial: acid etching was found to increase $T_{\rm c}$
\cite{RemeikaPL67,WanlassJSSC75} and it was
not clear if it was due to a reduction of the Rb content in the bulk
or only at the surface of the samples.\cite{KingLT1373,BevoloPRB74} 
These rather
confusing results even led Lefkowitz\cite{LefkowitzF77} to attribute 
any higher $T_{\rm c}$'s observed in non-stoichiometric samples to surface
effects. We
have undertaken to resume the experimental investigations on Rb$_x$WO$_3$
also encouraged by the acknowledging that interesting phenomena are
often hidden behind poorly reproducible data.

A great part of our work had to deal with physical chemistry issues. 
It will be described in a forthcoming paper,\cite{Brusetti} 
hereafter referred to as II.

\section{\label{experimental}Experimental}

\subsection{\label{preparationsamples}Preparation of samples}

In the first stage of our investigations, we studied single crystals
grown electrolytically from a melt consisting of Rb$_2$CO$_3$ and
WO$_3$---according to the method developed by Sienko and Morehouse.
\cite{SienkoIC63} 
Contrary to some authors who claimed to have obtained big single
crystals with $0.19<x<0.33$, we had good results, by this method,
only for the stoichiometric ($x\approx 0.33$) melt.  Moreover, it was very
difficult to guarantee that these crystals were completely free from
occlusions of the melt.  As our aim was, first, to clarify how the
superconducting properties depend on $x$, we needed homogeneous samples
and a good knowledge of their composition.  Therefore we preferred
carrying out this study with powder samples which have been prepared
by the usual solid-state reaction: for each nominal $x$ value, the
starting material was made up from high-purity Rb$_2$WO$_4$, WO$_3$ and W which
were ground together and placed in a quartz tube.  Before the tube was
sealed, the mixture was pumped to $10^{-6}$~Torr and baked repeatedly at
about 150$^\circ$C until we observed no more outgassing.  The tube was then
heated at 900$^\circ$C for 2 days.

As discussed in more details in II, we found no benefit from increasing 
the temperature above
900$^\circ$C, as we did not try to obtain single crystals; on the contrary
we observed that higher temperatures resulted in attack on the quartz
by the rubidium escaping from the Rb-rich samples.

\noindent This procedure yields fine crystalline powders with grain sizes
ranging between about 10 and 100~$\mu$m (bigger crystals being found near
the $x\approx 0.33$ content).  By checking that the X-ray powder diffraction
patterns of these samples agree with the HTB symmetry, and display no
trace of another phase, we can be quite confident in taking for $x$ its
nominal value.  This has been confirmed by the micro-probe analysis of
some of the samples.

It should be noticed that a solid-state reaction cannot
lead to a perfectly homogeneous non-stoichiometric compound: each
crystal is poorly connected to its neighbors in a low-pressure vapor
and should reach an equilibrium slightly depending on its morphology
and neighborhood.  From our micro-probe analysis we estimate the
dispersion of the Rb content in our batches at $0.005\lesssim 
\delta x\lesssim 0.01$.

\subsection{\label{characterization}Characterization of the superconducting 
transition}

We used a mutual-inductance bridge and a standard $^4$He cryostat to
systematically characterize the superconducting transition by
monitoring the diamagnetic expulsion in a small ($<2\times 10^{-4}$~T) low
frequency (33~Hz) magnetic field.  Magnetic susceptibility
measurements in static fields have also be done with a SQUID
magnetometer.  These methods are better suited to the investigations 
we were concerned with than conductivity
measurements. The width
of the magnetic transition gives a good picture of the homogeneity of
the sample, whereas a zero-resistance state may be due only to minute
superconducting sheets or filaments.  Moreover, as we shall show
later, a crystal which was grown from the vapor and has a suitable
size for conductivity measurements may not be representative of the
nominal composition of the batch, whereas this can be checked readily
on powder samples by taking them from different parts of the powder
batch.

We also carried out magnetic susceptibility and conductivity
measurements on various single crystals but these experiments---which
will be discussed in due course---will not be included in our
determination of the $T_{\rm c}$ vs $x$ dependence.

The amplitude of the magnetic transition of the powder samples could
vary on a rather large range ($\pm 20$\%), but we determined that this was only
related to the size and morphology of the grains as well as to the
compactness of the samples.

\section{\label{results}Results}

We tried to find out the cause(s) of the non-reproducibility of the
superconducting transition temperature $T_{\rm c}$ by first looking for a phase
transition between room temperature and 900$^\circ$C. We used differential
thermal analysis (DTA), differential scanning calorimetry (DSC) and
tried to anneal and quench the samples from various temperatures but
we found no sign of a phase transition.  However, these heat
treatments have brought to light some aspects of the physical
chemistry of these compounds which could explain a part of the
discrepancies previously noticed.  We shall deal with this in II and
here describe what appeared to be the major cause of these
discrepancies: the $T_{\rm c}$ of the non-stoichiometric compounds is sensible
to the cooling rate below room temperature. This was the expected
consequence of an ordering of the Rb vacancies like the one described
by Sato \textit{et al.},\cite{SatoPRB82} but we first missed this cooling-rate effect because
we tried to detect it in the region of the composition range where the
previously observed dispersion of $T_{\rm c}(x)$ was the highest, i.e. near 
$x\approx 0.25$. In this range, we never observed superconductivity above 1~K,
contrary to what claimed Stanley \textit{et al.}\cite{StanleyPRB79}.

\subsection{\label{coolingrateeffect}The cooling rate effect}

We have observed this effect for $0.29\lesssim x\lesssim 0.31$ and 
for $0.19\lesssim x\lesssim 0.23$, and
it is exemplified in Fig.~\ref{Brusettifig1}:
\begin{figure}
\centering{\includegraphics[width=7cm]{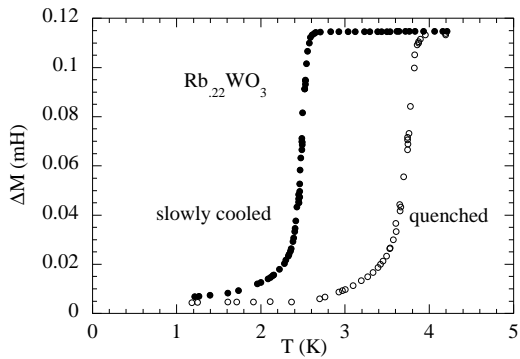}}
\caption{\label{Brusettifig1}Superconducting transitions 
(mutual inductance variations)
of a powder sample of Rb$_{0.22}$WO$_3$ after different coolings.}
\end{figure}
the ``slowly-cooled'' 
sample has been
cooled down to 90~K within a few hours before we transferred liquid
helium; the ``quenched'' same sample reached 4.2~K within a couple of
minutes after it has been introduced directly in the cryostat filled
with liquid helium.  As shown in Fig.~\ref{Brusettifig1}, it is the 
``slowly-cooled''
samples which have a lower $T_{\rm c}$, contrary to what is 
expected---and most usually observed---for a better ordered state. 
We also experimented with a faster cooling down to 90~K by dropping 
and stirring the sample in
liquid nitrogen but we observed no further increase of $T_{\rm c}$.  We also 
observed that the quenched state can be annealed at temperatures above
$T_{\rm m}\approx 110$~K$\pm 10$~K and, in this respect, we saw no difference between
the three samples studied with $x=0.19$, $x=0.22$ or $x=0.29$.

The cooling rate effect clearly confirms the ordering of the Rb atoms
in the non-stoichiometric HTB and their extreme mobility. We shall present 
now the other information we got on the order-disorder transformation.

\subsection{\label{calorimetricstudy}Calorimetric study of the order-disorder 
transformation}

\begin{figure}
\centering{\includegraphics[width=6.15cm]{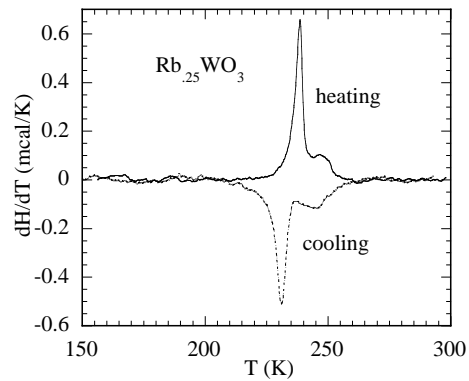}}
\caption{\label{Brusettifig2}DSC thermograms observed on a powder sample of 
Rb$_{0.25}$WO$_3$ ($\approx 50$~mg) on heating and on cooling at
$\pm 10$~K/min.}
\end{figure}

\begin{figure}
\centering{\includegraphics[width=7cm]{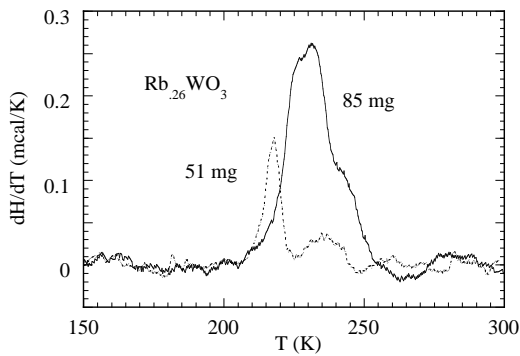}}
\caption{\label{Brusettifig3}DSC thermograms observed
on heating (10~K/min) for two powder samples of Rb$_{0.26}$WO$_3$.}
\end{figure}

After several unsuccessful attempts, we finally observed the enthalpy
anomaly accompanying this transformation, but only in the samples with
$x\approx 0.25$.  The apparatus we used was a Perkin-Elmer DSC-7 differential
scanning calorimeter and our measurements extended down to $\sim 100$~K. As
the anomaly is quite weak, we had to use rather sizeable powder
samples and high heating or cooling rates.  This could partly explain
the large temperature range of the anomaly and the thermal hysteresis
which are exemplified in Fig.~\ref{Brusettifig2}.
However, by comparing the
behaviors of different samples, it appears that the width of the
anomaly is mainly related to its strong $x$ dependence and to the
imperfect homogeneity of the samples: the data presented in 
Fig.~\ref{Brusettifig3}
indicate that a composition difference $\delta x\approx 0.01$ in the
neighborhood of $x=0.25$ leads to a shift of the anomaly by about 
20~K---the largest extrapolated peak onset temperature we observed being
about 240~K.

Our results are in rather good agreement with those of Sato 
\textit{et al.},\cite{SatoPRB82}
however, we observed no anomaly for $0.16\lesssim x\lesssim 0.23$ and 
$0.27\lesssim x\lesssim 0.33$ which
shows that the transition is first-order only for $x$ in a narrow range
around $x=0.25$.  Outside this range, the transition should become
continuous and, actually, it is what is observed in the
powder-diffraction pattern of Rb$_{0.27}$WO$_3$.

In our $x\approx 0.25$ samples, we estimate the maximum heat 
absorption $\Delta H$ which accompany the order-disorder 
transition to be $80\pm 10$~J/mol~K, and the maximum entropy increase 
$\Delta S$ to be $0.35\pm 0.05$~J/mol~K.  This value indicates
that the degree of order in Rb$_{0.25}$WO$_3$ is quite low just before it
transforms into the high-temperature disordered state.  Actually, the
molar configurational entropy of the disordered state is then
associated with the number of ways ($w$) of arranging $N$ rubidium
vacancies over $4N$ sites, i.e. $S_{\rm dis}=k_{\rm B}\ln w$, with :
\begin{eqnarray*}
w=\frac{N!}{\left(\frac{N}{4}\right)!\left(\frac{3N}{4}\right)!}
\quad\mbox{and}\quad N= N_0/3,
\end{eqnarray*}
as $x=0.25$ corresponds to 1/4 of the Rb atoms missing on the $N_0/3$
sites per mole available in the HTB structure.  It gives $S_{\rm 
dis}=1.56$~J/mol~K. If we eliminate all the arrangements where two vacancies or
more are first neighbors we find that $w$ should be:
\begin{eqnarray*}
w=\frac{\frac{3N}{4}!}{\left(\frac{N}{4}\right)!\left(\frac{N}{2}!\right)}
,\quad\mbox{which gives }S_{\rm dis}=1.32~\mbox{J/mol~K}.
\end{eqnarray*}
Accordingly, the increase of entropy $\Delta S$ we observed at the first-order
transition amounts to about 22--26\% of $S_{\rm dis}$.  This is in rather good
agreement with the simple Bragg-Williams approximation applied to AB$_3$
alloys, which gives $\Delta S/S_{\rm dis}=17.5$\%.

\subsection{\label{resistivityanomaly}The resistivity anomaly}

As mentioned earlier, Sato \textit{et al.}\cite{SatoPRB82} had noticed that the
order-disorder transition temperature $T_s$ they determined from their
neutron diffraction data seemed to correspond to the temperature 
$T_{\rm B}$ of
the resistivity anomaly observed by Stanley \textit{et 
al.}.\cite{StanleyPRB79} It was therefore
very tempting to attribute this anomaly to the ordering of the Rb
atoms. Since our results clearly confirmed this ordering and showed its
strong effect on the stability of the superconducting state, it was
still more difficult to understand why this resistivity anomaly had
not been observed by other authors.  This led us to undertake new
resistivity measurements on single crystals obtained by following the
same procedure as Stanley \textit{et al.},\cite{StanleyPRB79} i.e., 
a solid state-reaction at 
950$^\circ$C for 5 days.  We prepared in this way about 5~g of 
Rb$_{0.26}$WO$_3$ and
obtained the usual poorly crystallized, partly sintered powder, plus a
few mg of needle-like crystals. These often radiate in bundles from
tiny crystals attached to the quartz surface; this shows that they
grew from the vapor. Most of these needles were dark
blue and displayed the hexagonal symmetry of the HTB, but we also observed 
light-blue whiskers and even transparent ones, some of which 
were curved and even spiraled. We selected some of the dark blue 
needles, which were about 5~mm long in the $c$ axis direction and rather 
ribbon-shaped, therefore well suited to resistivity measurements 
(about 20--50~$\mu$m in width and 2--5~$\mu$m in thickness). 

Silver or gold paints, which are usually used to attach current and
potential leads to small samples, do not give satisfactory electrical
contacts on Rb$_x$WO$_3$, even by using freshly prepared crystals. 
While evaporated coatings are not very effective, we
obtained very good results by first sputtering gold on the contact
areas.  This
non-conducting behavior of the sample surface will be considered
in II.  The device we used produces contact areas partly
enveloping the samples, specially at their ends, which should allow
fairly uniform current lines.

We measured the temperature dependence of the resistivity of seven of
these samples between 300~K and 1.2~K, by using always slow cooling or
heating rates (about 2~K/mn) and we observed:
\begin{itemize}
\item[-] in two samples (A), an anomalous hump in the resistivity 
($\rho$)
as a function of temperature (Fig.~\ref{Brusettifig4})
\begin{figure}
\centering{\includegraphics[width=7cm]{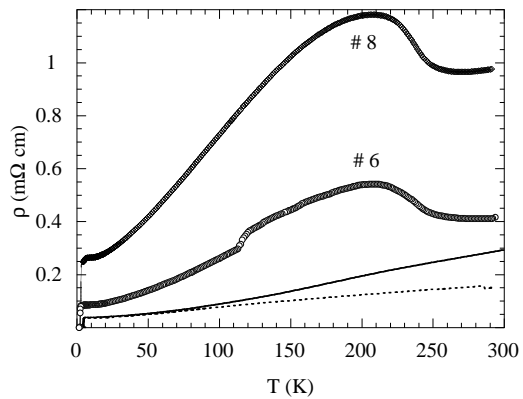}}
\caption{\label{Brusettifig4}Temperature dependence of the resistivity 
($\parallel c$) of two vapor-grown type-A fibers, compared with two 
massive acid-etched samples ($x\approx 0.19$) cut ($\perp c$) in an
electrochemically-grown crystal (full and dashed curves).}
\end{figure}
---similar to the one
observed by Stanley \textit{et al.},\cite{StanleyPRB79}
\item[-] in the other samples (B), a
low-temperature upturn of $\rho$ (Fig.~\ref{Brusettifig5}),
\begin{figure}[t]
\centering{\includegraphics[width=7cm]{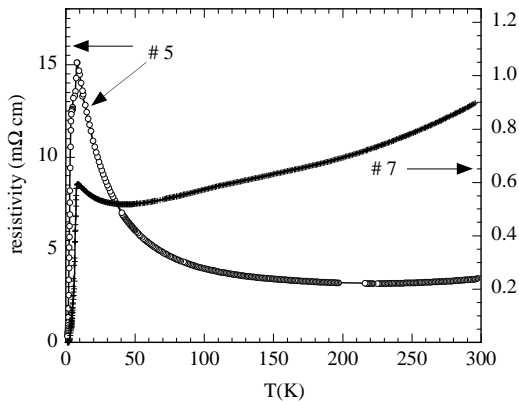}}
\caption{\label{Brusettifig5}Temperature dependence of the resistivity 
($\parallel c$) of two vapor-grown type-B fibers.}
\end{figure}
\item[-] in all samples, a
vanishing of $\rho$ occurring between $\approx 8$~K and $\approx 2$~K, 
(Fig.~\ref{Brusettifig6})
\begin{figure}[htb]
\centering{\includegraphics[width=7cm]{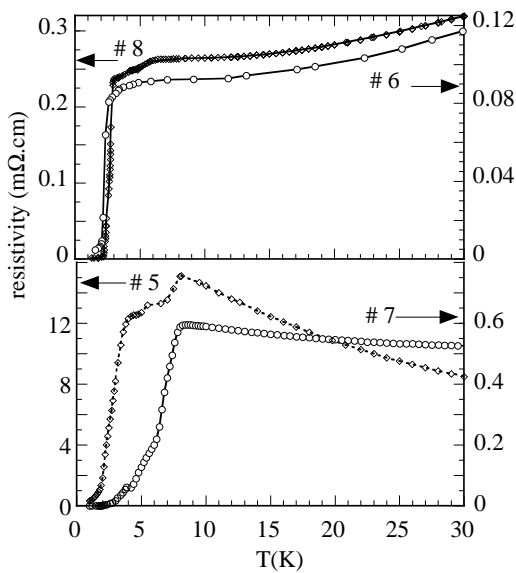}}
\caption{\label{Brusettifig6}Temperature dependence of the resistivity 
($\parallel c$) of the two types of vapor-grown fibers near the superconducting
transition.}
\end{figure}
whereas,
as usual, no sign of superconductivity was detected in the magnetic
susceptibility of the powder samples of the batch, 
\item[-] and large non-linear effects, in the same temperature range (see 
Sec.~\ref{nlresistivity}).
\end{itemize}
These conflicting observations prompted us to question the
identity of these fibers and to carry out their examination with a
JEOL 840-A scanning electron microscope (SEM) fitted with a energy
dispersive spectrometer (EDS).  This revealed that none of these 
vapor-transported crystals are really representative of 
Rb$_{.26}$WO$_3$: their rubidium content are much higher than the 
nominal content of the batch, samples B being structurally and 
chemically more homogeneous. We suspect
that the special transport properties of these fibers are consequences
of their structural non-homogeneity which could affect microscopic as
well as mesoscopic scales.  We suggest that crystallization from the
vapor might lead to the growth of rather decoupled layers of HTB with
sometimes, possibly, really different Rb contents.  Besides, each
layer or strip might be often perturbed by extended defects.  We shall
discuss the possible nature of these interlayer and interlayer
defects in II.

Within this framework we propose the following description of the
features listed above.

\noindent - The main decrease of the resistance---occurring at about 2.5~K in
samples A---is probably associated with the superconducting transition
of nearly stoichiometric Rb$_{0.33}$WO$_3$ strips.  But in some samples, and
particularly in samples B, the resistance begins to decrease at about
8~K, which is much higher than the highest $T_{\rm c}$'s ever observed in HTB
via magnetic measurements (cf.~Sec.~\ref{Tc(x)phasediagram}).  We think 
we are not dealing here with intrinsic superconductivity but rather 
with superficial or filamentary effects similar to those observed 
in NbSe$_3$,\cite{KawabataSSC82} for instance.

\noindent - The low temperature non-linear conductivity is a symptom which is
related to the defects mentioned above.

\subsection{\label{nlresistivity}The non-linear resistivity behavior 
of the vapor-grown fibers}

\begin{figure}[htb]
\centering{\includegraphics[width=7cm]{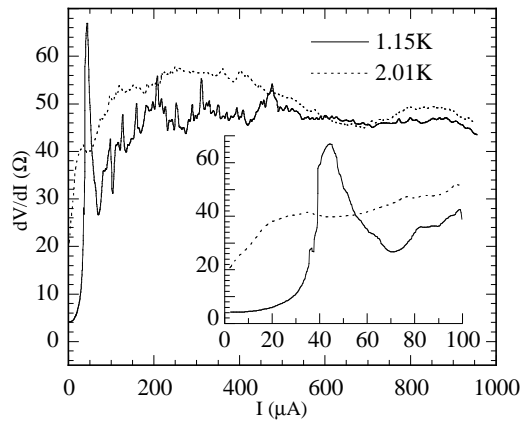}}
\caption{\label{Brusettifig7}Low-temperature
differential resistance of the vapor-grown sample $\# 5$ as a function
of the bias current. The inset magnifies the low-current range.}
\end{figure}

\begin{figure}[htb]
\centering{\includegraphics[width=7cm]{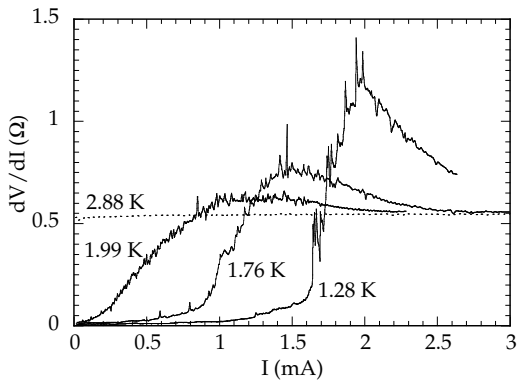}}
\caption{\label{Brusettifig8}Low-temperature differential resistance 
of the vapor-grown sample $\# 6$ as a function of the bias current.}
\end{figure}

At temperatures below about 2~K, the differential resistance 
${\rm d}V/{\rm d}i$
steeply rises when the bias current exceeds a critical value 
$I_{\rm c}$ and
then decreases toward the value of the normal state resistance
(Figs.~\ref{Brusettifig7}
and \ref{Brusettifig8}).
Sweeping 
the bias current up or down at different
rates does not modify significantly the current dependence of 
${\rm d}V/{\rm d}i$ which is not very sensitive either to 
thermal cycling.  Well below and also beyond $I_{\rm c}$ the 
differential resistance displays rather erratic
features which seem to be related to noise generating instabilities.

Assuming a uniform current distribution, the $I_{\rm c}$ values 
at $\approx 1.2$~K
would correspond to current densities $j_{\rm c}$ ranging between about 
0.05~A/mm$^2$ and 20~A/mm$^2$, depending on the samples.  We found 
no correlation
between this spread of the $j_{\rm c}$ values and the morphology 
of the samples
or their apparent resistivities---which are also widely spread and
significantly greater than quoted in the literature: at 300~K they
range between $\approx 4\times 10^{-4}~\Omega\,$cm and $3.5\times 
10^{-3}~\Omega\,$cm 
whereas the published
values stand between $3\times 10^{-5}~\Omega\,$cm and $1.5\times 
10^{-4}~\Omega\,$cm.

We shall notice
also that non-ohmic behaviors are observed too, above 8~K, in the
``normal'' state of those samples which display the greatest resistivity
upturn: there, at temperatures up to $\approx 20$~K, increasing the bias
current leads to a significant decrease of the resistance.

We think that these non-linear effects originate from a distribution
of weak links throughout the samples which leads to complex tunnelling
effects: in particular, below $\approx 2$~K, the features exemplified in
Figs.~\ref{Brusettifig7} and \ref{Brusettifig8} are quite 
reminiscent of those observed in granular
materials and attributed to Josephson tunnelling between the
grains.\cite{GubserPB81}  This picture is consistent with the high resistivity of
these crystals and suggests that normal tunnelling could be
responsible for the non-ohmicity displayed in the normal state below 
about 20~K.

We shall now sum up the information we have on the anomalous hump in
resistivity which motivated this study: we only observed this hump in
two of seven vapor-grown samples and never observed it---whatever
the Rb content---in the more massive crystals prepared by fused-salt
electrolysis.\cite{wehave} In these the resistivity decreases quite
linearly with temperature down to about 100 K, before it progressively
saturates (Fig.~\ref{Brusettifig4}). This is the behavior also observed by
Aristim\'uno \textit{et al.}\cite{AristimunoJSSC80} in crystals which 
were carefully selected to be electrically homogeneous. On the contrary, 
we showed that the
vapor-grown crystals are not homogeneous and display a transition
towards a zero-resistance state which begins at about 8~K. This 
resistance anomaly is not observed magnetically in any bulk HTB. The 
foregoing remarks lead us to suspect that the
resistivity anomaly is extrinsic and related to the inhomogeneity of
the samples. However, its occurrence in the same temperature range as
the order-disorder transition is certainly not merely a matter of
chance.  Actually the same Florida State University group observed the
same anomalies in K$_x$WO$_3$,\cite{CadwellPRB81} but at higher 
temperatures, and no anomaly
in Cs$_x$WO$_3$\cite{SkokanPRB79}---which is consistent with 
the greater tendency for the
smaller ions to order.  As described below, this also agrees with the
quite different $T_{\rm c}(x)$ dependences observed in the three bronzes. 
The
only explanation we can offer to rationalize these behaviors is the
following: the ordering of the Rb atoms leads to a reduction of the
density of states at the Fermi level $\cal{D}_{\rm FE}$ which 
destabilizes the superconducting state but has only a minute 
effect on the resistivity
of the normal state in the massive samples.\cite{GubserBrusetti} 
On the other hand, in
the vapor-grown samples, the loss of conductivity associated with the
structural defects---those revealed by the non-linear effects described
above---could be enhanced by the reduction of $\cal{D}_{\rm FE}$ and 
lead to the hump observed.

\subsection{\label{Tc(x)phasediagram}The $T_{\rm c}(x)$ phase diagram}

\begin{figure}[b]
\centering{\includegraphics[width=7cm]{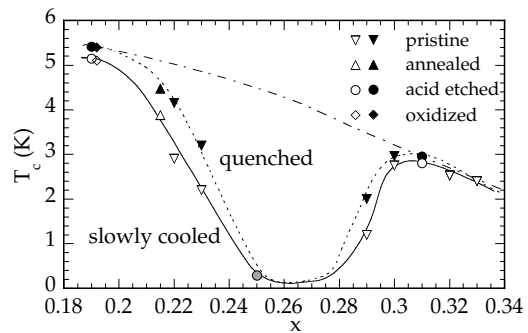}}
\caption{\label{Brusettifig9}The superconducting transition 
$T_{\rm c}$ as a function of the rubidium content $x$. 
Empty or full markers refer to measurements after slow cooling or 
quenching from $\approx 300$~K respectively. The grey marker 
corresponds to an intermediary cooling in a $^3$He-$^4$He dilution 
refrigerator. The curves are only guides to the eye and the 
monotonous one extrapolates what we think would be the $T_c(x)$ 
dependence if the vacancy ordering could be prevented.}
\end{figure}

The information collected above
confirms that we should not rely on the composition of the
vapor-grown samples to establish the tungsten bronzes $T_{\rm c}(x)$ phase
diagram.  It also allows us to dismiss the high values of $T_{\rm c}$ observed by
Stanley \textit{et al.}\cite{StanleyPRB79} near $x$~= 0.25. As we just saw, 
they are probably to be
ascribed to a kind of filamentary superconductivity, whereas the $T_{\rm c}$'s
of the corresponding bulk material deeply decrease---as a result of the
ordering of the Rb vacancies.  The $x$ dependence of $T_{\rm c}$ therefore
displays a pronounced dip near this Rb content which allows the
greatest degree of order and leads to the highest transformation
temperature $T_{\rm ord}$ (Fig.~\ref{Brusettifig9}).
This dip is less 
pronounced when the samples are
quenched from room temperature but ordinary quenching rates are not
sufficient to prevent a significant ordering when $0.23\lesssim 
x\lesssim 0.28$, and
superconductivity cannot develops above $\approx 1$~K within this composition
range. Beyond this range, $T_{\rm c}$ steeply increases with $x$, up to 
$T_{\rm c}\approx 3$~K
for $x\approx 0.30$, and then slowly decrease down to $T_{\rm 
c}\approx 2$~K, when the Rb
content approaches the stoichiometric value.  On the other side of the
dip, $T_{\rm c}$ increases still more abruptly with the decrease of the Rb
content and seems to level off at $T_{\rm c}\approx 5$~K for 
$x\lesssim 0.21$.  However, this
asymptotic behavior seems to conflict with the higher $T_{\rm 
c}$'s ($\approx 5.5$~K)
we can obtain in samples in which the Rb content has been reduced by
acid etching or by a slight oxidation.\cite{itwasknown}

\begin{figure}[htb]
\centering{\includegraphics[width=7cm]{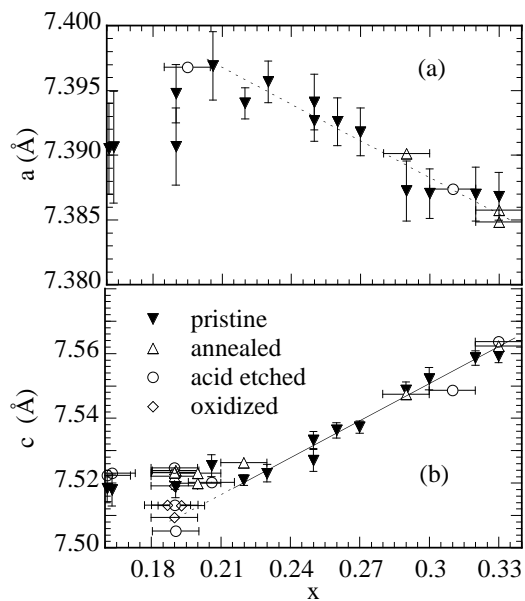}}
\caption{\label{Brusettifig10}Dependence of the lattice parameters 
of the HTB phase as a function of the rubidium content. 
Nominal $x$ values are used for the pristine samples whereas the 
mean values deduced from the SEM analysis are used for the other 
samples. Below $x\approx 0.20$, most of the data for the a parameter 
have been omitted because of their very poor accuracy---probably 
due to the blurring effect of the parasitic phase.}
\end{figure}

This inconsistency prompted us to question the stability of the HTB
phase for $x\lesssim 0.22$. By examining more systematically 
the X-ray powder patterns, we found that the $x$ dependence of the
lattice parameters shows a discontinuity near $x\approx 0.215\pm 0.005$ 
in samples prepared by the usual solid-state reaction method 
(Fig.~\ref{Brusettifig10}).
This clearly indicates that another phase 
coexists with the HTB's when the rubidium content falls below this 
value. In the slowly
cooled samples, the corresponding diffraction peaks begin to appear
unambiguously only when $x\lesssim 0.19$, a value which was therefore associated,
up to now, with the boundary of the HTB-phase region. In these samples,
we found it is WO$_3$ which coexists with the HTB phase and gives hardly
perceptible diffraction peaks---which agree with the fact that 
Rb$_{0.21}$WO$_3$
should be in equilibrium with only about 10\% of WO$_3$ when the nominal
Rb/W ratio is 0.19. In samples which were quenched at
the end of their otherwise similar preparative treatment, the larger additional peaks
can be attributed to the ITB phases and seem compatible with a content
amounting to 20\%.  These observations are consistent with an 
ITB-WO$_3$ boundary situated near 700$^\circ$C--800$^\circ$C, as 
proposed by A.~Hussain.\cite{HussainACSA78}

On the other hand, the acid-etched or slightly oxidized samples in
which we observed the highest $T_{\rm c}$'s, display also the lowest
$c$-parameter values.  Moreover, these values correspond to what is
expected for an HTB phase with $0.19\lesssim x\lesssim 0.20$---extending 
the linear decrease
of $c$ with $x$ observed at higher Rb contents (see 
Fig.~\ref{Brusettifig10}).\cite{thecorresponding} 
Correspondingly, we think that the superconducting transitions which
begin above 5~K in the latter samples are to be attributed to this
structure and rubidium content.

This led us to complete the low-$x$ region of $T_{\rm c}(x)$
diagram as displayed in Fig.~\ref{Brusettifig9}. The transition 
temperature $T_{\rm c}$ has
been defined here at the onset of the diamagnetic expulsion, as the
width of the transition considerably depends on the samples: it can be
only about 0.2~K in nearly stoichiometric samples ($x\lesssim 0.30$) which are
less sensitive to the ordering transformation, but it can
be more than 1~K when the Rb content of the samples leads to the
stronger cooling-rate effect.  We also observed that powder samples
with $x\lesssim 0.23$, which were annealed at 950$^\circ$C and partly sintered,
displayed much steeper transitions, but they recover their pristine
behavior when they are powdered again.  This probably indicates that
the penetration depth $\lambda$ in these samples is close to the size of a
notable proportion of the grains, i.e. of the order of 1~$\mu$m at least.

Although at first sight it looks quite different, this $T_{\rm c}(x)$ diagram
cannot but recall the corresponding Cs$_x$WO$_3$ diagram which displays a
monotonic increase of $T_{\rm c}$ with decreasing $x$.  Actually, the strong
reduction of $T_{\rm c}$ near $x\approx 0.25$ in Rb$_x$WO$_3$ is the signature of the
rubidium vacancies ordering. This effect is still more pronounced in
K$_x$WO$_3$---where the ordering is very easy\cite{KrauseSSC88}---whereas 
it is absent in the
cesium tungsten bronze, where no noticeable ordering seems to occur. 
In other words, the $T_{\rm c}(x)$ diagrams of the potassium and rubidium
hexagonal tungsten bronzes would be quite similar to the diagram of
Cs$_x$WO$_3$, if we could prevent the smaller alkali ions from ordering. 
Finally, we shall notice that the $x$ dependences we obtained for
the lattice parameters in Rb$_x$WO$_3$---as displayed in 
Fig.~\ref{Brusettifig10}---are now
quite similar to the corresponding ones in Cs$_x$WO$_3$
\cite{HussainACSA78}: the decrease of
the alkali-metal content leads to a decrease of the $c/a$ ratio; the
larger alkali metals giving the larger effect and, probably, playing a
more effective role in the stabilization of the HTB phase, as
indicated by the fact that they lead to a farther extended homogeneity
range, on its low-$x$ side.

\section{\label{discussion}Discussion}

\subsection{\label{Tc(x)issue}The $T_{\rm c}(x)$ issue}

We shall first remark that this increase of $T_{\rm c}$
with decreasing x was also observed in the tetragonal tungsten bronzes
(TTB) Na$_x$WO$_3$ and K$_x$WO$_3$.\cite{Ngai78} Well before 
the discovery of the high $T_{\rm c}$
cuprates this behavior aroused a great interest.  The point is that
it seems to conflict with the simple BCS model because a decrease of
the alkali content should correspond to a decrease of the density of
state at the Fermi level $\cal{D}_{\rm FE}$. This has been well established in the
case of the most extensively studied tungsten bronze
Na$_x$WO$_3$.\cite{SienkoACS63,experimental} The increase of $T_c$ 
with decreasing $x$ in these compounds is therefore probably due to 
an enhancement of the electron-electron interaction in the low-$x$ 
range. Although we have less clues to support this diagnosis as far as 
Rb$_x$WO$_3$ is concerned,\cite{note} we think it remains the most 
likely in view of the many analogies between the two systems.

Salchow \textit{et al.}\cite{SalchowJPCS83} attributed the increase 
of $T_c$ with increasing $x$ to a better screening of the 
electron-phonon interaction. We think that this mechanism is probably 
not very effective in the HTB, because it would imply that the 
ordering of the alkali vacancies---which certainly reduces 
$\mathcal{D}_{\rm FE}$---would increase $T_c$, whereas it is 
drastically reduced. Another mediation of the electron-electron 
interaction has been proposed by Kahn and Ruvalds\cite{KahnPRB79} who 
invoked acoustic plasmons but, at the moment, we consider that the 
reliable experimental data on these bronzes are in favour of those 
models which rely on the lattice instability.

Actually the crystal structure of these bronzes is not very stable: 
according to Sato \textit{et al.},\cite{SatoPRB82} a distortion of the 
WO$_3$ cage takes place in Rb$_x$WO$_3$ below about 420~K. Above this 
transition the space group is P6$_3$/$mcm$, but below this 
temperature the exact structure remained unsolved at that time. A 
similar poor agreement with any structural model is observed in 
K$_x$WO$_3$.\cite{KrauseSSC88,PyeMRB79}. Most of these studies 
indicate that the topology of the oxygen octahedra is disturbed 
around an alkali vacancy. Moreover, the great versatility of these 
octahedra is well known: they are able to join along different 
directions by sharing their corners, or even their edges, and to yield 
to various deformations---which makes the structures they form able 
to sustain large deviations from stoichiometry. This leads to the 
spectacular ``crystallographic shear'' (CS) planes in WO$_{3-x}$, to 
the lamellar intergrowth of HTB and WO$_3$ in low-alkali-content 
``intergrowth tungsten bronzes'' (ITB) and to many WO$_6$-based 
metastable structures formed at rather low temperatures, \textit{via} 
the methods of soft chemistry.

On the other hand, we have many evidences of the great mobility of 
the small alkali atoms in the tungsten bronzes. In the HTB, 
particularly, low-frequency Einstein-like phonon modes are 
associated to the vibrations of these atoms within the large channels 
running along the c-direction.\cite{ChesserF77} These modes were made 
responsible for the ``excess'' heat capacity observed above about 
10~K, and which can be fitted---in Rb$_{.33}$WO$_3$---by an Einstein 
contribution with $\Theta_E\approx 60$~K.\cite{BevoloPRB74} However, 
as their characteristic energy seems to hardly depend on 
$x$,\cite{SatoPRB82} they probably are not the main cause for the 
anomalous $T_c(x)$ dependence.

Other special phonons have been invoked, originally by 
Shanks,\cite{ShanksSSC74} and have given rise to the LSE model 
proposed by Ngai \textit{et al.},\cite{NgaiPRB76} then treated by 
Vuji\v{c}i\'c \textit{et al.}\cite{VujicicJPC81}. They attribute the 
enhancement of the electron-electron attractive interaction, in the 
tungsten bronzes, to local structural instabilities: their model is 
based on the idea that a local ground-state configuration could be 
separated from one or more other configurations by a small amount of 
energy and by a potential barrier. According to these authors, the 
excitations between these local states---they called ``local 
structural excitations'' (LSEs)---could not only enhance the phonon 
pairing of electrons but also mediate a supplementary effective 
electron-electron interaction through a specific electron-LSE 
interaction.

One of the strong points of this model lies in the fact that the 
local structural instabilities are associated with the alkali 
vacancies and it is conceivable that increasing their content should 
increase the instability of the lattice and the part of the LSE in its 
dynamics, whereas the ordering of these vacancies should have the 
opposite effect. Following the approach of Ngai \textit{et 
al.},\cite{NgaiPRB76} we may imagine, in the case of Rb$_x$WO$_3$, 
that there are microscopic regions where the local density $x_\ell$ is 
smaller than the average value $x$ and thus where there is a tendency, 
for the WO$_6$ octahedra, to collapse into a WO$_3$ structure---which 
reminds us of the intergrowth mechanism leading to the ITB phases 
when $x\leq 0.2$. Where a local transformations has occurred, a kind 
of WO$_3$ defect has been formed. On the other hand, if the local 
transformation has not condensed, the local structural instability is 
preserved, and the elemental WO$_3$ defect can be viewed as an excited 
state of the local group of atoms and bonds. Finally, it is also 
tempting to speculate that the large amplitude vibrations of the 
alkali atoms could facilitate the tunneling between the free-energy 
minima of two structural configurations or, more generally, that 
combinations of these local instabilities with the alkali-atom 
vibrations could lead to a specially effective coupling with the 
conduction electrons.

The LSE model has been adopted by Sato \textit{et al.}\cite{SatoPRB82} 
and applied to Rb$_x$WO$_3$ by using the McMillan's 
equation\cite{McMillanPR68} where they introduced an electron-phonon 
coupling parameter $\lambda_{\rm LSE}$ proportional to $n_{\rm 
LSE}$---the number of LSE--- and to $\mathcal{D}_{\rm FE}$. This 
simple approach is able to describe the $x$ dependence of $T_c$ as it 
provides the necessary increase of $\lambda$ with the decrease of $x$. 
However the inelastic neutron scattering experiments undertaken by 
these authors on powder samples have not allowed to observe any 
significant $x$ dependence. As proposed by these authors, a tentative 
explanation of this failure could lie in the broadness of the LSE 
spectrum; however that may be, this issue should deserve to be 
experimentally settled.

\subsection{\label{interfacialsupercon}Interfacial superconductivity}

Besides this problematic $T_{\rm c}(x)$
dependence, superconductivity in Rb$_x$WO$_3$ displays another unusual
feature which has been revealed by our resistivity measurements on the
vapor-transported samples, namely its stabilization---at temperatures
above the highest $T_{\rm c}$'s of the bulk material---in probably interfacial
regions.  Such a phenomenon has been observed in WO$_{3-x}$,
\cite{AirdJPL98} and attributed
to twin walls.  Sheet superconductivity also develops on the surface
of WO$_3$ crystals which have been subjected to a slight superficial
enrichment of sodium.\cite{ReichEPJB99} The very high $T_{\rm c}$'s 
observed in the later case
(up to 91~K) is evidently far from being explained but indicates how
much the interfacial properties of these materials could be promising. 
The great versatility of the WO$_6$ octahedra we discussed above,
certainly plays a part in these superficial or interfacial properties.

An explanation of such enhancements of $T_{\rm c}$ was proposed, twenty five
years ago, by Lefkowitz,\cite{LefkowitzF77} who thought that 
the anomalous $T_{\rm c}(x)$
dependence in the tungsten bron\-zes was a surface effect: he stressed
that a ferroelectric instability can condense at low temperatures in
WO$_3$ and hypothesized that this could lead to high electric fields at
the boundary between the insulating material and the doped
regions---this inducing a new electron density of states at the Fermi
level.  Although we are now quite sure that the increase of $T_{\rm c}$ with
the reduction of the alkali content, in these bronzes, is really a
bulk property, the Lefkowitz's proposition seems fairly seductive when
considering the phenomenon we observed in the vapor-transported
samples.

\subsection{\label{odtransformation}The order-disorder transformation}

Finally we shall add a few
comments about the order-disorder transition of the rubidium vacancies
in Rb$_x$WO$_3$.  It was observed first by neutron diffraction 
measurements\cite{SatoPRB82}
and then by electron diffraction measurements.\cite{Monfort82}  Our calorimetric
measurements have shown that this transformation was first-order only
for $x\approx 0.25$, which corresponds to one forth of the Rb atoms missing,
to the most stable ordering and to a doubling of the lattice
constants.  It appears that the positions of the O atoms are modulated
to some extent by the alkali vacancies,\cite{SchultzSato} therefore a gap at the Fermi
surface is certainly induced by this new periodicity.  Subsequently,
the K ordering in K$_x$WO$_3$ was also studied\cite{KrauseACB85} and 
different ordering
schemes have been proposed. A quite complex picture emerge from these
studies, which indicates, one more time, that we are dealing with an
``infinitely adaptive structure''.\cite{KihlborgCS88}

When we move away from $x\approx 0.25$,
the transition temperature $T_{\rm ord}$ rapidly decreases: it is only about
200~K in Rb$_{0.22}$WO$_3$\cite{SatoPRB82} and about 123~K in 
Rb$_{0.20}$WO$_3$\cite{Monfort82} i.e., very near from
our estimate of the rubidium mobility threshold $T_{\rm m}$, and this agrees
with our observation that the disordered state, then, can be quenched
more effectively.  Conversely, the low-temperature state will be
usually less ordered.  Moreover, the complex ordering taking place
when $x<0.25$ has probably a weaker effect on the Fermi surface. 
Therefore, both mechanisms contribute to reduce the cooling-rate
effect when $x$ approaches its minimum value.

Let us finally notice
that similar issues, concerning order-disorder phenomena and their
influence on superconductivity, have been raised in the high-$T_{\rm c}$
cuprates.\cite{VealPRB90} A thermodynamical model for the ordering schemes of
the oxygen vacancies has been proposed,\cite{deFontaineJLCM90} which leads to a
concentration dependence of the ordering temperature quite reminiscent
of the $T_{\rm ord}(x)$ behavior observed in Rb$_x$WO$_3$; in particular, a
first-order transition is predicted only in a part of the
concentration range.  However, ordering increases $T_{\rm c}$ in these cuprates
as well as in the low-$T_{\rm c}$ organic superconductors, whereas it is just
the opposite in the HTB.

\section{\label{conclusion}Conclusion}

We have identified the main reasons why the available results on the
physical properties of the HTB M$_x$WO$_3$ were so contradictory.  We have
described here the effect of the order-disorder transition and
interface-related artifacts; other causes of discrepancies result
from the physical chemistry of these compounds and will be described
in the forthcoming paper II. We have now a much clearer vision of the
HTB, which brings to the fore a common unusual feature of the
superconducting state whose critical temperature $T_{\rm c}$ increases when the 
M content $x$ and $\cal{D}_{\rm FE}$
decrease.  This behavior is also shared by the tetragonal tungsten
bronze Na$_x$WO$_3$.  After reviewing the reliable available information on
these systems we concluded that a strong-coupling mechanism was
probably responsible for this feature.  At this respect, the model
appealing to local structural excitations (LSE) seems the most 
attractive, and it is backed up by our observations that the 
superconducting state is destabilized by the ordering of the M 
vacancies, whereas $T_{\rm c}$ increases with their number. Actually, it is 
the only model, up to now, able to conciliate these two features. 
However, we established that superconductivity is nearly completely 
suppressed in the HTB when 1/4 of the M atoms are missing and 
ordered, and this cannot be due only to the LSE mechanism as 
superconductivity exists up to $\sim 2$~K in the stoichiometric HTB. 
Thus, the M-atoms ordering also has a deep effect on the Fermi 
surface---in contradiction with the previous rigid-band descriptions. 
Conversely, this indicates that the strong-coupling LSE mechanism--- 
if present---must be the most effective when $T_{\rm c}$ is the highest 
and $\cal{D}_{\rm FE}$ is simultaneously reduced by the highest 
vacancy content and by their ordering, i.e. for $x$~= 0.19. It is 
therefore near this concentration that more 
extensive lattice-dynamic studies should be carried out. More generally, 
from a fundamental 
point of view and irrespective of the quest for room-temperature 
superconductivity, we think that the possible building up of an 
attractive electron-electron coupling \textit{via} localized 
interactions should be investigated most seriously and that 
Rb$_x$WO$_3$ is now a system where this task may be tackled 
profitably.

\acknowledgments
We are grateful to A. Sulpice for his magnetization
measurements and to Y. Monfort for communication of unpublished data. 
We greatly acknowledge stimulating discussions with O. B\'ethoux and the
assistance of P. Amiot and A. Hadj-Azzem in the X-ray powder
diffraction measurements and SEM examinations and analysis.


\begin{thebibliography}{99}
\bibitem{wehave}
We have prepared some non-stoichiometric crystals by annealing 
stoichiometric electrocrystallized crystals within a large sample of 
compacted non-stoichiometric powder, or by acid etching (see II).
\bibitem{itwasknown}
It was known for a long time that acid etching could remove the Rb 
ions from the HTB and considerably enhance $T_{\rm c}$.
\cite{RemeikaPL67,WanlassJSSC75,KingLT1373} 
We have investigated this effect and we propose a description of the 
chemical process involved in II. There we also describe how a 
moderate oxidation, at about 450$^\circ$C, can have a similar effect 
on the sample core, by driving the Rb ions towards the superficial 
tungstate layer.
\bibitem{thecorresponding}
The corresponding increase of the $a$ parameter is more difficult to 
bring to the fore because about four times smaller. Moreover, in the 
oxidized samples it is somewhat obscured by the presence of tungstate 
diffraction peaks.
\bibitem{experimental}
Experimental evidences (specific heat and photoemission spectroscopy) 
and band structure calculations are discussed in 
Ref.~\onlinecite{NgaiPRB76} for instance.
\bibitem{note}
In the case of Rb$_x$WO$_3$, we have only rough
estimates of $\cal{D}_{\rm FE}$: one can be derived from the Pauli-Peierls-type
contribution to the magnetic susceptibility\cite{WanlassJSSC75} 
i.e., $\cal{D}_{\rm FE}$ $\approx 0.3$~state/eV.mol---with no sensible 
variation with $x$---and another is
provided by the thermodynamical properties of the superconducting
state\cite{BevoloPRB74} i.e., $\cal{D}_{\rm FE}$ $\approx 0.7$~state/eV.mol 
in Rb$_{0.33}$WO$_3$. These
estimates are quite similar to the values obtained in Na$_x$WO$_3$ and, in
both cases, are compatible with what can be expected from a
nearly-free-electron band, accommodating the electrons donated by the
alkali atoms.
\end{thebibliography}
\end{document}